\documentstyle[aps,amssymb,epsf,preprint]{revtex}
\newcommand{\beq}{\begin{equation}}
\newcommand{\eeq}{\end{equation}}
\newcommand{\bea}{\begin{eqnarray}}
\newcommand{\eea}{\end{eqnarray}}
\begin{document}

\title{ Threshold criterion for wetting at the triple point}

\author{
S. Curtarolo$^{1,3}$, G. Stan$^1$, M. J. Bojan$^2$, M. W. Cole$^{1,4}$, and W. A. Steele$^2$} 
\address{ $^1$Department of Physics, Penn State University, University Park, PA 16802, USA\\
$^2$Department of Chemistry, Penn State University, University Park, PA 16802, USA\\
$^3$Present address: Dept. of Materials Science and Engineering, MIT, Cambridge, MA 02139, USA\\
$^4${corresponding author, e-mail: mwc@psu.edu, phone: (814) 863-0165, fax: (814) 865-3604}
}
\date{\today}

\maketitle
\begin{abstract}{\sf
Grand canonical simulations are used to calculate adsorption isotherms of
various classical gases on alkali metal and Mg surfaces. {\em Ab initio}
adsorption potentials and Lennard-Jones gas-gas interactions are used.
Depending on the system, the resulting behavior can be nonwetting for all
temperatures studied, complete wetting, or (in the intermediate case) exhibit 
a wetting transition. An unusual variety of wetting transitions at the triple
point is found in the case of a specific adsorption potential of
intermediate strength. The  general threshold for wetting near the triple
point  is found to be close to  that predicted with a heuristic model of
Cheng {\em et al.} This same conclusion was drawn in a recent experimental and
simulation study of Ar on CO$_2$ by Mistura {\em et al.} These results imply 
that a dimensionless wetting parameter $w$ is useful for predicting whether
wetting behavior is present at and above the triple temperature. The
nonwetting/wetting crossover value found here is $w \simeq$ 3.3.
}

{\it PACS numbers: 64.70.Fx, 68.35.Rh, 68.45.Gd, 82.20.Wt}
\end{abstract}
\newpage

\section{Introduction}                   

Many adsorption systems exhibit either complete wetting below the triple
point or triple point wetting transitions. These behaviors are recognized
to arise from strongly attracting substrate potentials and/or small
mismatch of substrate/adsorbate lattice constants \cite{ref.3.pandit,ref.3.degennes}.  
Less attractive potentials exhibit incomplete or nonwetting behavior at the 
triple point. There is considerable interest in the nature of such weakly 
attractive situations near and above this temperature \cite{ref.3.cole}. General 
arguments and model calculations of Cahn \cite{ref.3.cahn} and of Ebner and 
Saam \cite{ref.3.ebner} in 1977 implied that systems which are nonwetting at the 
triple point ought to exhibit transitions to wetting at some temperature below the
critical point. Experimental confirmation of this proposal was not found until
appropriate systems were identified. In 1991, it was argued that alkali metals
provide the weakest adsorption potentials of any surfaces for He atoms and 
therefore wetting transitions were expected for such substrates at tow temperature
\cite{ref.3.cheng1}. 
This hypothesis was subsequently confirmed and the corresponding transition 
has since been studied in many laboratories 
\cite{ref.3.nacher,ref.3.rutledge,ref.3.ketola}. 
More recently, similar phenomena have been predicted and/or seen for H$_2$, 
Ne, and Hg on various surfaces 
\cite{ref.3.cheng2,ref.3.ross1,ref.3.mistura1,ref.3.hess,ref.3.hensel,ref.3.kozhevnikov}. 
Key theoretical questions arising from these studies involve the 
reliability of both the assumed adsorption potentials and the statistical 
mechanical description of the transition, as well as the effects of surface
heterogeneity \cite{ref.3.curtarolo} present in any real experiment.

In the course of these investigations, Cheng {\em et al.} \cite{ref.3.cheng1}
(CCST henceforth)
posited a simple model which interprets the transitions in terms of a
balance between the surface tension cost of producing a thicker film and the
energy gain associated with the film's interaction with the surface, V(z)
\cite{ref.3.cheng3}. There results an implicit equation for the wetting temperature
(T$_w$):
\begin{equation}
(\rho_l - \rho_v) I_V \ = \ 2 \gamma
\end{equation}
\begin{equation}
I_V \ = \ - \ \int_{z_{min}}^\infty dz\ V(z) 
\label{label.3.eq:integral}
\end{equation}
where $\rho_l$ and $\rho_v$ are the number densities of the adsorbate
liquid and vapor at coexistence, $\gamma$ is the surface tension of the
liquid, and $z_{min}$ is the equilibrium distance of the potential.
If one assumes a simple 3-9 form of the adsorption potential
\begin{equation}
 V= (4 C^3)/ (27 D^2 z^9) - C/ z^3
\end{equation}
then the transition condition becomes
\begin{equation}
 (C D^2)^{1/3} = 3.33 \gamma/(\rho_l - \rho_v)
\label{label.3.eq:criterion}
\end{equation}
CCST used this model to predict wetting transitions for a large number of
specific systems. More recently, the same general approach was used by
Chizmeshya {\em et al.} (CCZ) \cite{ref.3.chizmeshya}, who based their predictions 
on new {\em ab initio} adsorption potentials. The latter potentials differ from
 earlier potentials in several respects; especially important are differences 
in values of the coefficient C of the van der Waals asymptotic potential, 
resulting in significantly larger well depths D. There is indication that 
these revised potentials are more consistent with experimental data than are 
their predecessors. Particularly striking is the improved agreement for $^4$He
contact angle \cite{ref.3.ross2,ref.3.klier,ref.3.ancilotto1} and wetting temperature on Cs, the 
most studied wetting transition to date \cite{ref.3.nacher,ref.3.rutledge,ref.3.ketola}.

These predictions have also been tested by recent experiments of 
Hess {\em et al.} \cite{ref.3.hess} and simulations of Ne on alkali metals and 
Mg \cite{ref.3.bojan}. In this paper, we extend the simulation studies to a 
larger family of systems for which the hypothesis is relevant. We focus here 
on the following  question: assuming that V(z) is known, does the CCST model 
correctly predict whether a specific adsorption system exhibits wetting at and 
above the triple temperature? 

The outline of this paper is the following. Section \ref{label.3.sec:method} summarizes
our simulation methodology  and presents detailed results for several systems
which yield fairly distinct adsorption behaviors. These include complete
wetting for all T (Kr/Mg and Xe/Mg), an anomalous wetting transition near the
triple point (Kr/PMg), a prewetting transition line (Ar/Mg) somewhat above
the triple temperature, and prewetting transitions characterized by the 
formation of very thick films (for Ar, Kr and Xe on Li).

Here PMg is our own nomenclature for ``pseudo-magnesium'', an ersatz surface
which attracts Kr atoms about 7\% more strongly than Mg itself 
\cite{ref.3.pseudoMg}. We also evaluate other hypothetical substrates in order to 
ascertain the triple point wetting threshold.
In section \ref{label.3.sec:interpretation}, we compare these findings with those 
obtained previously in simulations and/or experiments for inert gases on 
various surfaces \cite{ref.3.sukhatme,ref.3.mistura2,ref.3.bruschi,ref.3.krim}.

\section{Method and Results}             
\label{label.3.sec:method}

The simulation technique has been described in detail previously \cite{ref.3.bojan},
so only a brief description is given here. For this study, isotherms for 
various model systems were calculated using the Grand Canonical Monte 
Carlo (GCMC) technique \cite{ref.3.frenkel,ref.3.allen}. For each isotherm, 6.8 million 
steps (each step being an attempted creation, deletion, or displacement of a 
molecule) were performed to reach equilibration and 4.5 million steps were 
performed in the data gathering phase. For some points near the 
transition regions, where the fluctuations are larger, the number of 
equilibration steps was  increased to 20 million and 12 million steps were 
performed for data gathering. 

As in our previous work, the model for the atom-atom interaction is a 
Lennard-Jones 12-6 potential and the atom surface interaction is modeled 
using the potential described by  Chizmeshya {\em et al.} \cite{ref.3.chizmeshya}.
These potentials can be characterized by four parameters for
each system studied. A well-depth $\epsilon_{gg}$ and distance
$\sigma_{gg}$ are needed to specify the adsorbate LJ potential
and a well-depth $D$ and the van der Waals constant $C$
specify the atom-surface potential. The values of these
parameters used in this study are listed in Table \ref{label.3.table:param}. 
The parameters correspond to Ne, Kr, Ar, and Xe adsorbed on
alkali and alkali earth metal surfaces. 
The surface labeled PMg (pseudo magnesium) 
was obtained when we attempted to simulate 
Xe on Li, but used the 
$\epsilon_{gg}$ and $\sigma_{gg}$ values for Kr instead of Xe. This potential
is similar to that of Kr on a Mg surface; however, the Kr/PMg 
distance of closest approach is 20\% smaller and the well depth 
is about 7\% weaker than estimated for  Kr on Mg. While this potential does not 
correspond to a ``real'' system, we shall see that this small decrease in 
interaction strength significantly affects the wetting properties of the 
substrate/adsorbate system.

In most cases the simulation cell had a height of h = 70 \AA, 
with the metal surface 
forming the boundary at one end and a hard wall surface at the 
other. In Xe/Li and Ar/Li simulations we increased the height to 140 \AA\,and
210 \AA.
The surface unit cell was taken to be $10\ \sigma_{gg} \times 10\ 
\sigma_{gg}$ giving a nominal monolayer coverage of roughly 100 atoms. Periodic
boundary conditions in these dimensions render the surface effectively
infinite; we used a fairly large cutoff (5$\sigma_{gg}$) to minimize long 
range corrections. 

For each system, we simulated adsorption isotherms over
relevant  temperature ranges from the triple point to the critical temperature.
The results are shown in figures \ref{label.3.fig:XeLi1}-\ref{label.3.fig:KrMg}. 
Of the systems studied the least attractive surface 
of the surface interactions are for the Ar/Li, Kr/Li and Xe/Li cases. 
In all of these, we see evidence of prewetting transitions and,
based on our results, we report wetting 
temperature. 
For Ar/Li we estimate T$_w = 130 \pm4$ K, for Kr/Li,
T$_w = 175\pm4$ K and T$_w=225 \pm 4$ K for Xe/Li.
All $T_w$ are reported in Table \ref{label.3.table:param}.

Despite the fact that these systems are examples of prewetting transitions,
there are some qualitative differences from the case (say Ne/Mg)
where the wetting transition appears close to the triple temperature.
Figure \ref{label.3.fig:XeLi1} shows a series of isotherms for Xe/Li 
from $T^*=$ 1.0 to 1.25.
The isotherm at 221 K (the lowest $T^*$) 
shows nonwetting behavior, manifested as slight adsorption below
saturated vapor pressure. The higher temperature isotherms however, show 
prewetting transitions below saturated vapor pressure, from which we estimate $T_w$. 
The nature of the prewetting jump can be further characterized by the results
in figures \ref{label.3.fig:XeLi2}, \ref{label.3.fig:XeLi3},
and \ref{label.3.fig:XeLi4}, all at T=254 K. 
Figure \ref{label.3.fig:XeLi2} shows the dependence of
the isotherm on the height $h$ [$=70, 140, 210$ \AA] of the simulation box. 
The rapid rise occurs at the same pressure, independent of box height. 
This is a clear indication that the transition is prewetting rather than
capillary condensation. The detailed
behavior is most easily understood by looking at the
density profiles' dependences on $h$. Figure \ref{label.3.fig:XeLi3} 
shows these profiles at $P=22.9$ atm.  
Note that the data for two largest boxes coincide but differ from
those of the smallest box. Evidently, this regime requires quite large
values of $h$ in order to obtain reliable data; this is not surprising
because the compressibility of the system is particularly high at this
transition as the slope of the isotherms suggest. 
The same phenomenon is even more dramatic for the case of a 1\%
higher pressure. As seen in figure \ref{label.3.fig:XeLi4}, 
one must go to even larger $h$ ( $>200$
\AA) in order to derive accurate adsorption data even though the liquid
regime occupies only the region $z<70$ \AA. While this sensitivity to $h$
and the details of the transition are quite interesting they are not the
focus of this paper. Suffice it to say that the density jumps
discontinuously near $P=23.0$ atm, as seen in the different densities in
figures \ref{label.3.fig:XeLi3} and \ref{label.3.fig:XeLi4}.

We note one further point about these data which is that the film's density
close to the surface is qualitatively similar to that on a relatively more
attractive surface. There appears a quite large accumulation of adsorbate atoms
in the region associated with the first layer. There is even a significant density
maximum occurring where a dense second layer might occur. The latter is,
however, preempted by the prewetting jump in the regime of temperature seen
here. Above the prewetting critical point, in contrast, one does expect
such smooth film growth.

As the relative strength ($D/\epsilon_{gg}$) of the surface interaction increases, 
(Ar on Mg, Kr on PMg) the isotherms give evidence of wetting transitions. The 
results for Ar on Mg, (shown in figure \ref{label.3.fig:ArMg}) indicate 
prewetting transitions at about 10\% above the triple point (T$_{tr}$ = 81 K). 
From these data,  we estimate the wetting temperature to be 90 K for Ar/Mg 
with a prewetting critical temperature of 95 K. 

In figure \ref{label.3.fig:KrPMg}, the results for Kr on PMg are shown. 
This system shows clear prewetting transitions below 128 K for all 
temperatures above the triple temperature (estimated to be 116 K 
\cite{ref.3.Hansen}). 
Although this is a fictitious system, it represents one of the first
simulations of triple point wetting. The only other simulation of 
wetting in this region is the case of Ne/Mg which we reported in 
\cite{ref.3.bojan}.

Finally, as the surface interaction strength further increases, one begins to
see continuous growth for all T between the triple temperature and the
critical point. This is evident in figure \ref{label.3.fig:KrMg} 
which shows isotherms for Kr on Mg. The 
isotherms were simulated at temperatures near the estimated triple temperature 
and demonstrate complete wetting on the Mg 
surface by the adsorbate even at the lowest temperatures. 
The results for Xe on Mg (not shown) are qualitatively the same.

\section{Interpretation of the results}
\label{label.3.sec:interpretation}

To interpret our results we follow CCST, who evaluated their simple model in
terms of the law of corresponding states. Then we note that equation 
\ref{label.3.eq:criterion} may be written in a universal form in terms of the reduced 
well depth $D^*$ and interaction range $x$, defined as follows:
\begin{eqnarray}
D^* &=& D/\epsilon_{gg} \\
x &=& (C/ D)^{1/3} / \sigma_{gg}
\end{eqnarray}
The values of these parameters are reported in Table \ref{label.3.table:param}. 
The CCST model may then be written in terms of a dimensionless wetting parameter
\begin{equation}
w= x D^* =  3.33 \gamma/ [(\rho_l - \rho_v) \sigma_{gg} \epsilon_{gg}]
\end{equation}
The triple point experimental data summarized in CCST yield a value of 3.6
for the right side of this equation. Hence the criterion is $w$ = 3.6 for a
wetting transition precisely at the triple point.  The latter value,
however, presupposes a 3-9 adsorption potential, which we have not used in
the present simulations. If, instead, we compute the integral of the actual
potential used in the calculations, we find a value which is 8 to 
10\% higher than the result obtained with the 3-9 potential. Taking this
correction into account, the threshold for wetting at the triple point
becomes somewhat lower for the realistic potential:
\begin{equation}
w = 3.3
\end{equation}
This is shown in figure \ref{label.3.fig:threshold} as a dashed curve.
Turning to the actual simulation results, we find that this prediction
characterizes our data relatively well, insofar as the vicinity of the
value 3.3 is a crossover regime of wetting behavior. For all gases exposed
to the alkali metal surfaces the quantity $w$ is less than 2.3, and
the systems are nonwetting at the triple point. 
The case of Ar/Mg ($w$ = 3.2) is one for which a wetting transition
occurs about 10\% above the triple temperature; similar behavior was found
in our previous study of Ne/Mg ($w $= 3.4). The case of Kr/PMg involves a
similar value ($w$ = 3.2), but we find anomalous triple point wetting 
transition behavior for that system. Finally, the cases of Kr/Mg ($w$ = 3.4) 
and Xe/Mg ($w$ = 3.8) both correspond to complete wetting. For comparison with 
a more typical adsorption system, the case of Ar/Au ($w$ = 6.7) is one which 
exhibits wetting at and above the triple temperature \cite{ref.3.sukhatme,ref.3.krim}.

We may add to this list of relevant systems the recently studied case of
Ar/CO$_2$. Mistura {\em et al.} \cite{ref.3.mistura2} found that it exhibits 
triple point wetting; 
this finding agreed with their density functional calculation using their
theoretical potential. This potential would be roughly consistent with the CCST
model in that the latter would predict a wetting transition at a
temperature only 3\% above the triple point temperature, 83.8 K \cite{ref.3.Ar/CO2}.

While the criterion $w$ = 3.3 successfully discriminates between the
nonwetting and wetting data sets mentioned above, it is worthwhile to
ascertain  the validity of this criterion more accurately. Hence we have
performed extensive simulations at the triple temperature, varying the well
depth near the values $x$ = 0.7, 0.9, and 1.4 in order to determine the wetting
transition line. The results are shown in Figure \ref{label.3.fig:threshold}. One 
identifies this line as the point where the letter C is just above the letter 
W. Note that these transition values (dash-dot curve) lie somewhat above the 
$w$ = 3.3 dashed line. This comparison indicates that the CCST model 
underestimates the well depth criterion for wetting.

This finding is qualitatively consistent with  results reported recently by
Ancilotto and Toigo \cite{ref.3.ancilotto2}.
Using a density functional method, these workers found wetting transitions
very close to the coexistence curve for several systems (e.g. Ne/Li and Na)
for which our simulations find nonwetting behavior. The different findings of the simulations may be due to the uncertainty in identifying wetting behavior close to saturation, especially at T approaching T$_c$. In these and other
cases, the transitions occurred at (typically $\sim$ 10\%) higher temperature
than is predicted by the CCST model. These findings are also consistent
with one's realization that the CCST model predicts wetting transitions for
all adsorption circumstances, however weakly attractive, while a more reliable
theory yields drying transitions for {\em ultraweakly} adsorbing potentials
\cite{ref.3.ancilotto2}.

To summarize, we have found the CCST equation \ref{label.3.eq:integral} to be a 
semiquantitative guide to  the occurrence of wetting transitions. It would be 
valuable to have additional, genuine experimental data to test this conclusion.
 We wish especially to encourage experimentalists to explore totally new 
systems, especially Ar, Kr, and Xe adsorption on alkali metals. There are 
several reasons for doing so. One is that such experiments will provide tests 
of the weak adsorption potentials  for these systems, for which no experimental
 data exist. 
A second is that triple point wetting can involve interesting behavior 
at and just below the triple temperature, as has been found in many 
experiments \cite{ref.3.zimmerli,ref.3.T_triple}.

We have not explored either these systems, or analogous ones, below
the triple temperature. In order to reliably describe the crystalline
phase, as is necessary there, such simulations would require a more careful
handling of the unit cell dimensions and shape than has been employed here.
In view of the many relevant systems, such a study seems well justified for 
the future.

\acknowledgments

This research was supported by the National Science Foundation. It has 
benefited from discussion with F. Ancilotto, V. Bakaev, M. H. W. Chan, 
J. Z. Larese, G. Mistura, and F. Toigo. We are especially grateful to 
{\it Fondazione Aldo Ing. Gini} for the generous support of Ing. S. Curtarolo.

\begin{table}  [h]
\center
\caption{The values of the LJ parameters, $\epsilon_{gg}$ and $\sigma_{gg}$, 
for the gas-gas interactions and the values of the reduced well depth and width 
$D^*=D/\epsilon_{gg}$ and $x$, as defined in Eq. (6), 
for various systems studied and correlated with wetting behavior, 
defined as follows: C = continuous growth, TP = anomalous wetting transition
near or at the triple point (T$_{tr}$), W = wetting transition above the 
triple point, N = nonwetting for all T $>$ T$_{tr}$, and D = drying. 
The Ar/CO$_2$ data were taken from Ref. \protect \cite{ref.3.krim}. 
The wetting temperatures estimated by the simulations are in parentheses.}
\vspace*{0.2 in}
\begin{tabular} {c|ccccccc}
Gas / Surface		&	&Rb	&Na	&Li	&PMg	&Mg	&CO$_2$	\\\hline
	Ne		&$D^*$	&0.7	&1.1	&1.5	&	&2.8	&	\\
$\epsilon_{gg}=34$ K	&$x$	&1.6	&1.4	&1.4	&	&1.2	&	\\
$\sigma_{gg}=2.78$ \AA	&behavior&D	&N	&N	&	&TP(22K)	&	\\
			&$xD^*$	&1.1	&1.5	&2.1	&	&3.4	&	\\\hline
	Ar		&$D^*$	&1.1	&1.6	&2.1	&	&3.5	&3.8	\\
$\epsilon_{gg}=120$ K	&$x$	&1.1	&1.1	&1.0	&	&0.9	&1.0	\\
$\sigma_{gg}=3.41$ \AA	&behavior&N	&N	&W(130K)&	&W(90K)	&C	\\
			&$xD^*$	&1.2	&1.7	&2.1	&	&3.2	&3.8	\\\hline
	Kr		&$D^*$	&	&	&2.3	&3.6	&3.8	&	\\
$\epsilon_{gg}=171$ K	&$x$	&	&	&0.9	&0.9	&0.9	&	\\
$\sigma_{gg}=3.60$ \AA	&behavior&	&	&W(175K)&TP(116K)&C	&	\\
			&$xD^*$	&	&	&2.1	&3.2	&3.4	&	\\\hline
	Xe		&$D^*$	&	&	&2.8	&	&5.45	&	\\
$\epsilon_{gg}=221$ K	&$x$	&	&	&0.8	&	&0.7	&	\\
$\sigma_{gg}=4.10$ \AA	&behavior&	&	&W(225K)&	&C	&	\\
			&$xD^*$	&	&	&2.2	&	&3.8	&	\\
\end{tabular}
\label{label.3.table:param}
\end{table}

%%%%%%%%%%%%%%%%%%%%%%%%%%%%%%%

\begin{figure}%%%%%%%%%%%% FIGURE 1
\caption{Solid curves are adsorption isotherms for Xe/Li at 
T=221, 232, 243, 254, 265, 276 K. The vertical dashed lines to the right 
of each curve indicate the saturated vapor pressures.}
\label{label.3.fig:XeLi1}
\end{figure}

\begin{figure}%%%%%%%%%%%% FIGURE 2
\caption{Adsorption isotherms for Xe/Li at 
T=254 K as a function of simulation cell height. The heights for the curves
shown are 70 \AA (dashed), 140 \AA (dots), 210 \AA (full curve).
The vertical dashed line to the right indicates the saturated vapor pressure.}
\label{label.3.fig:XeLi2}
\end{figure}

\begin{figure}%%%%%%%%%%%% FIGURE 3
\caption{Film density for Xe/Li at T=254 K and P=22.9 atm as a function
of reduced distance. Curves correspond to varying heights of the simulation
cell: h=70 \AA (dots), 140 \AA (dashed), 210 \AA (dot-dashed), and 280 \AA
(full curve). Note the latter three results coincide.}
\label{label.3.fig:XeLi3}
\end{figure}

\begin{figure}%%%%%%%%%%%% FIGURE 4
\caption{Film density for Xe/Li at T=254 K and P=23.1 atm as a function
of reduced distance. Curves correspond to varying heights of the simulation
cell: h=70 \AA (dots), 140 \AA (dashed), 210 \AA (dot-dashed), and 280 \AA
(full curve). Note the latter two results coincide.}
\label{label.3.fig:XeLi4}
\end{figure}

\begin{figure}%%%%%%%%%%%% FIGURE 5
\caption{Adsorption of Ar on Mg for T = 89, 90, 91, 92, 94, 95, 96, and 98 K 
(from left to right). The data indicate a wetting transition near 90 K and a
prewetting critical point near 95 K.
The vertical dashed lines occur at saturated pressure for each temperature.}
\label{label.3.fig:ArMg}
\end{figure}

\begin{figure}%%%%%%%%%%%% FIGURE 6
\caption{Adsorption of Kr on ``pseudo-Mg'', an artificial problem described in 
the text. The isotherms (from left) begin at 119 K and run to 128 K. There is a
wetting transition close to the triple temperature, T$_{tr}$ = 116 K. The 
prewetting critical temperature is 127 K.}
\label{label.3.fig:KrPMg}
\end{figure}

\begin{figure}%%%%%%%%%%%% FIGURE 7
\caption{Adsorption of Kr on Mg, showing continuous growth of the film for all 
T above the triple temperature. Curves correspond to T = 120, 125, 130, 135,
140, and 145 K, from left to right.}
\label{label.3.fig:KrMg}
\end{figure}

\begin{figure}%%%%%%%%%%%% FIGURE 8
\caption{Points represent systems shown in Table \ref{label.3.table:param} or 
artificial systems studied as discussed in Section \ref{label.3.sec:interpretation}. 
Symbols are defined in the caption to this table. The threshold line based on 
the model discussed in the text, $w = D^* x = 3.3$, is indicated by dashes. The
 dash-dot curve represents the wetting threshold derived from our simulations.}
\label{label.3.fig:threshold}
\end{figure}

\newpage
\begin{figure}[ht]%%%%%%%%%%%% FIGURE 1
\epsfysize=5.5in \epsfbox{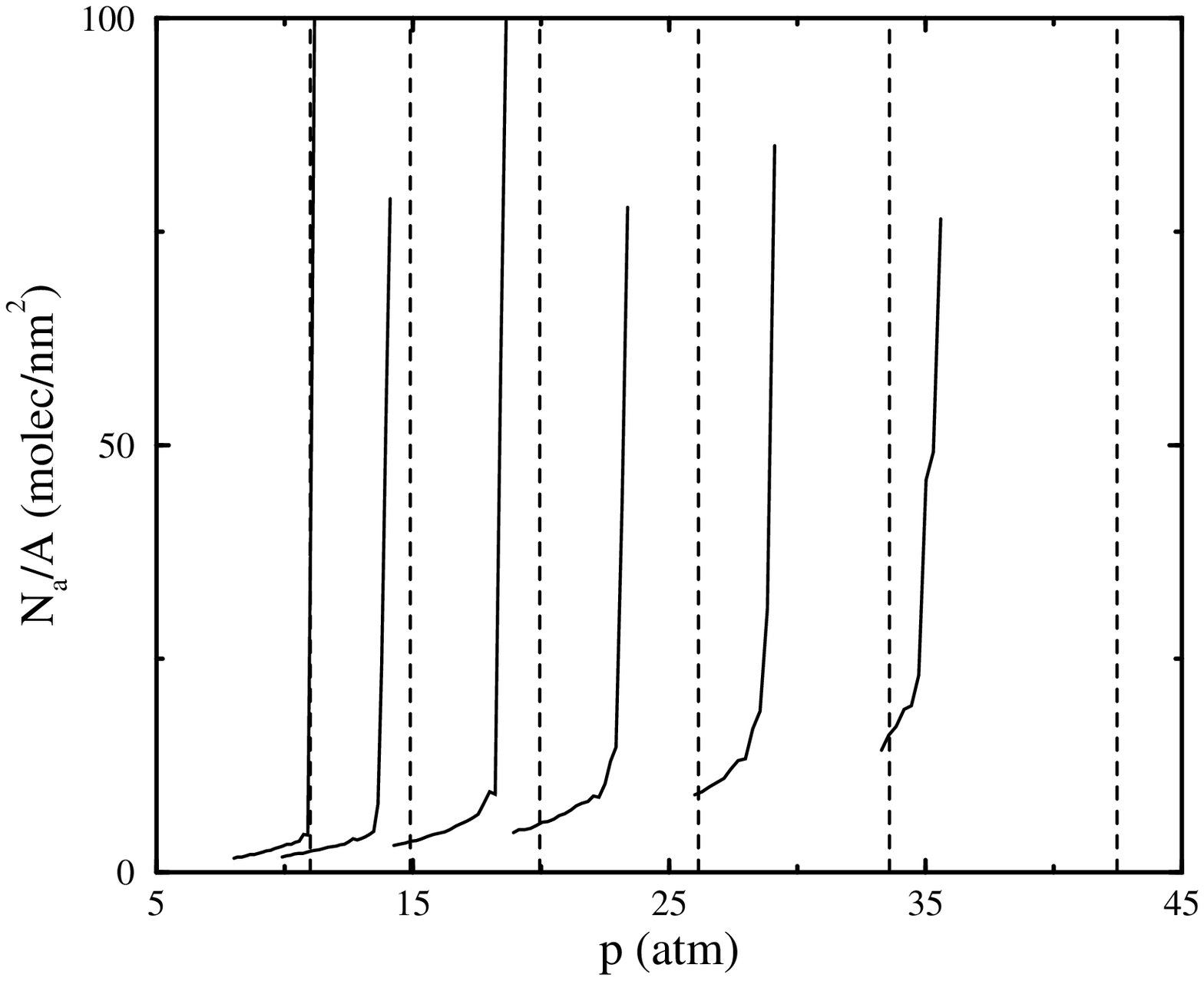}
\end{figure}
\vspace{4cm}\begin{center}{\bf FIG. 1}\end{center}

\newpage
\begin{figure}[ht]%%%%%%%%%%%% FIGURE 2
\epsfysize=5.5in \epsfbox{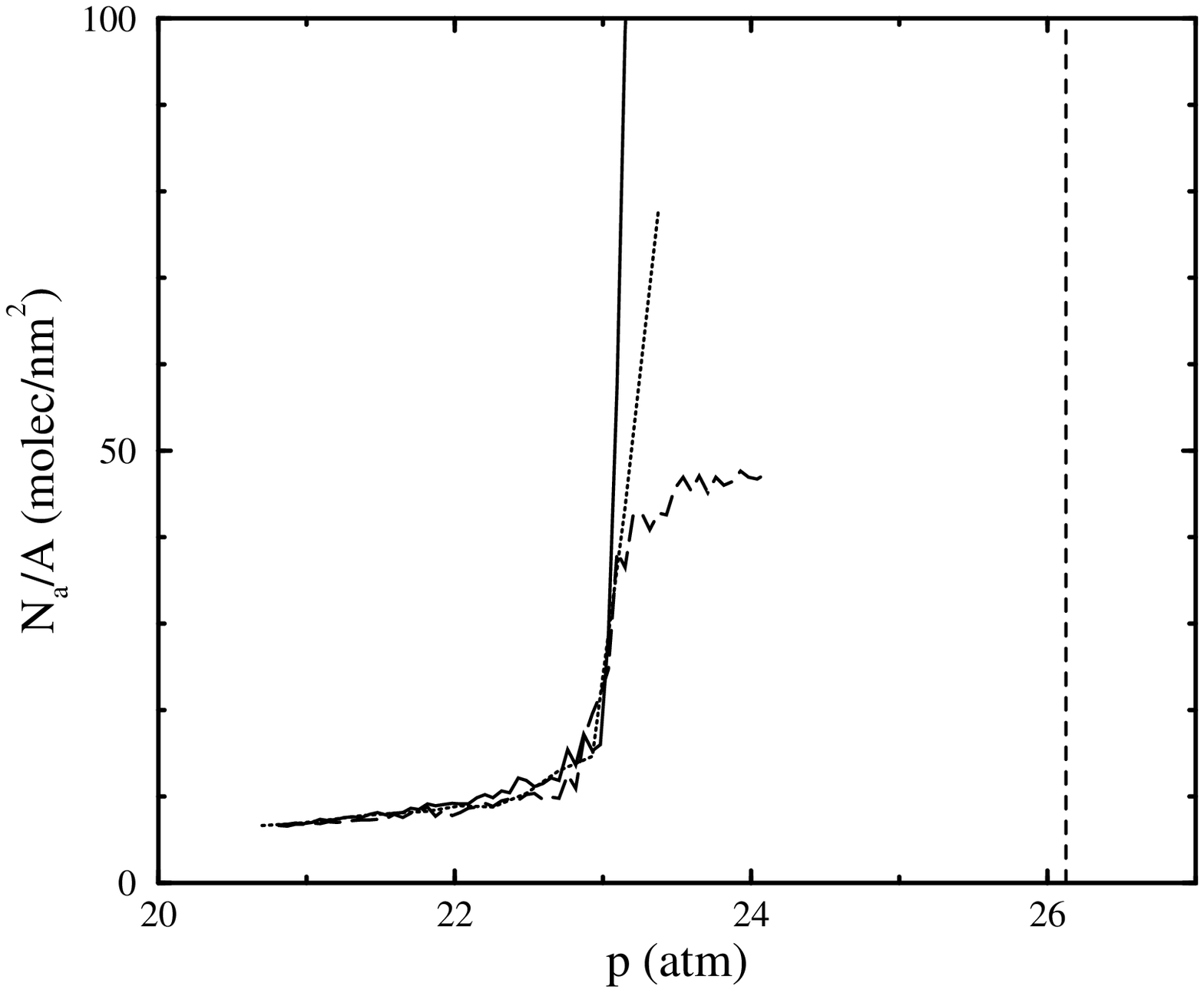}
\end{figure}
\vspace{4cm}\begin{center}{\bf FIG. 2}\end{center}

\newpage
\begin{figure}[ht]%%%%%%%%%%%% FIGURE 3
\epsfysize=5.5in \epsfbox{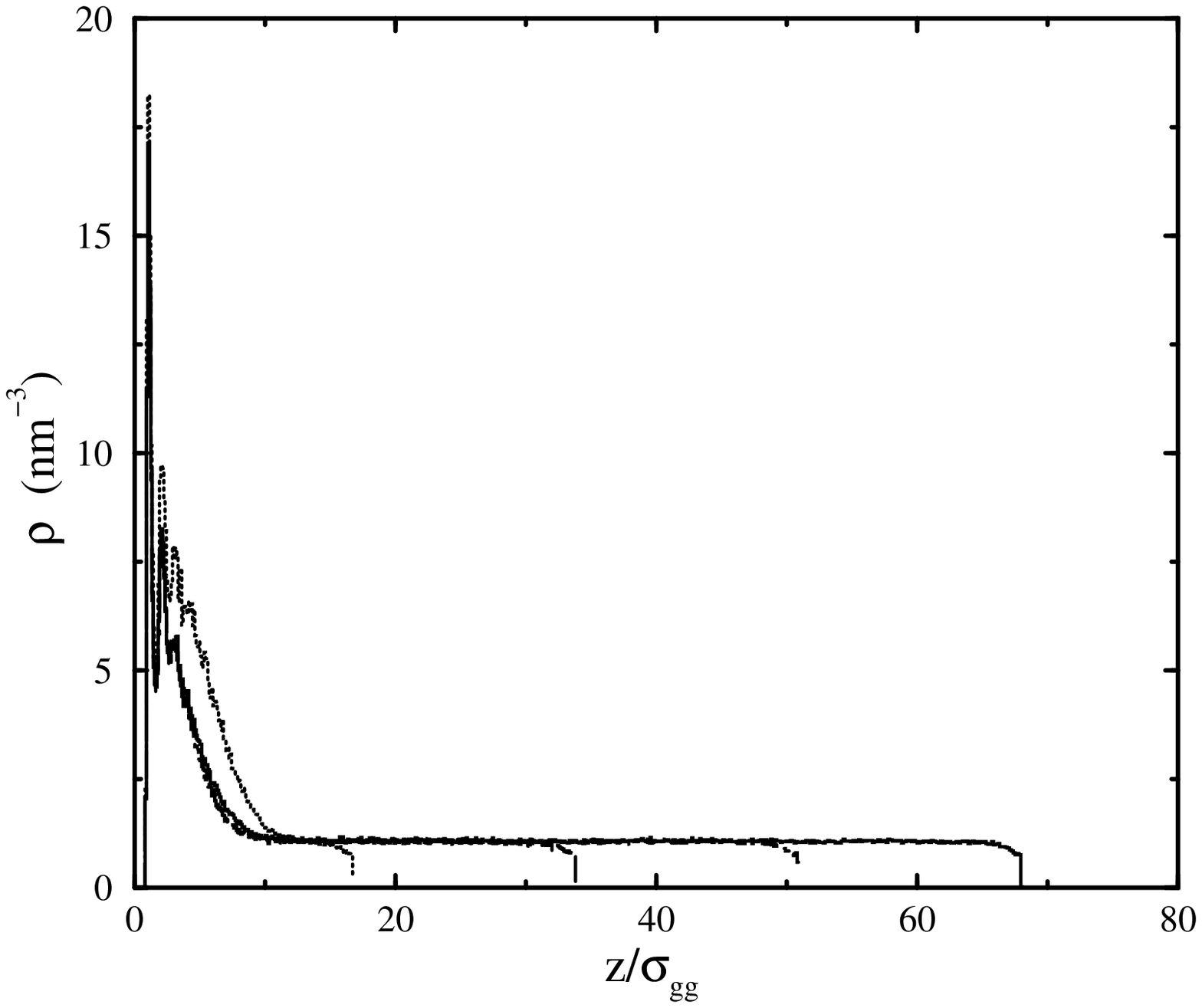}
\end{figure}
\vspace{4cm}\begin{center}{\bf FIG. 3}\end{center}

\newpage
\begin{figure}[ht]%%%%%%%%%%%% FIGURE 4
\epsfysize=5.5in \epsfbox{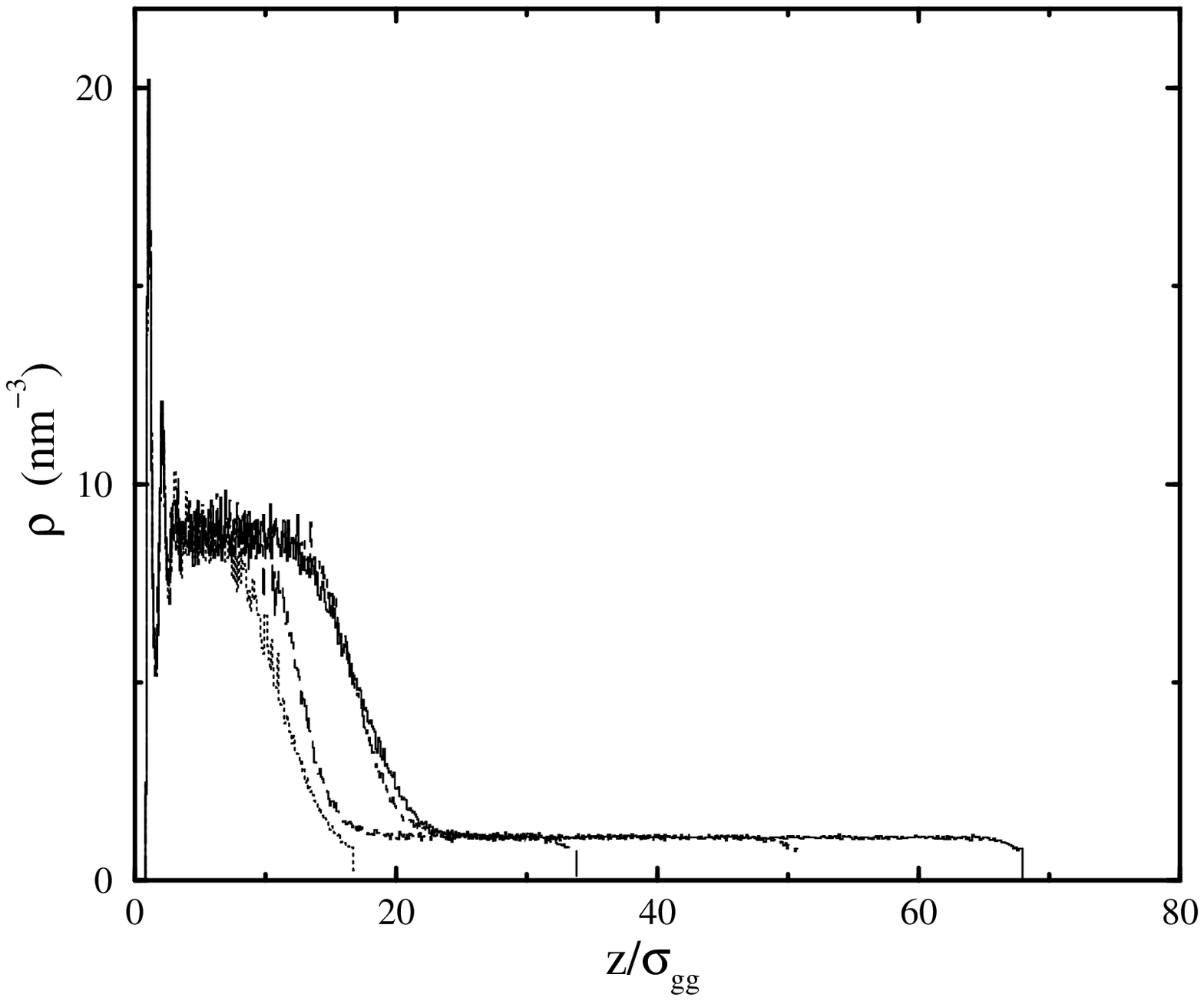}
\end{figure}
\vspace{4cm}\begin{center}{\bf FIG. 4}\end{center}

\newpage
\begin{figure}[ht]%%%%%%%%%%%% FIGURE 5
\epsfysize=5.in \epsfbox{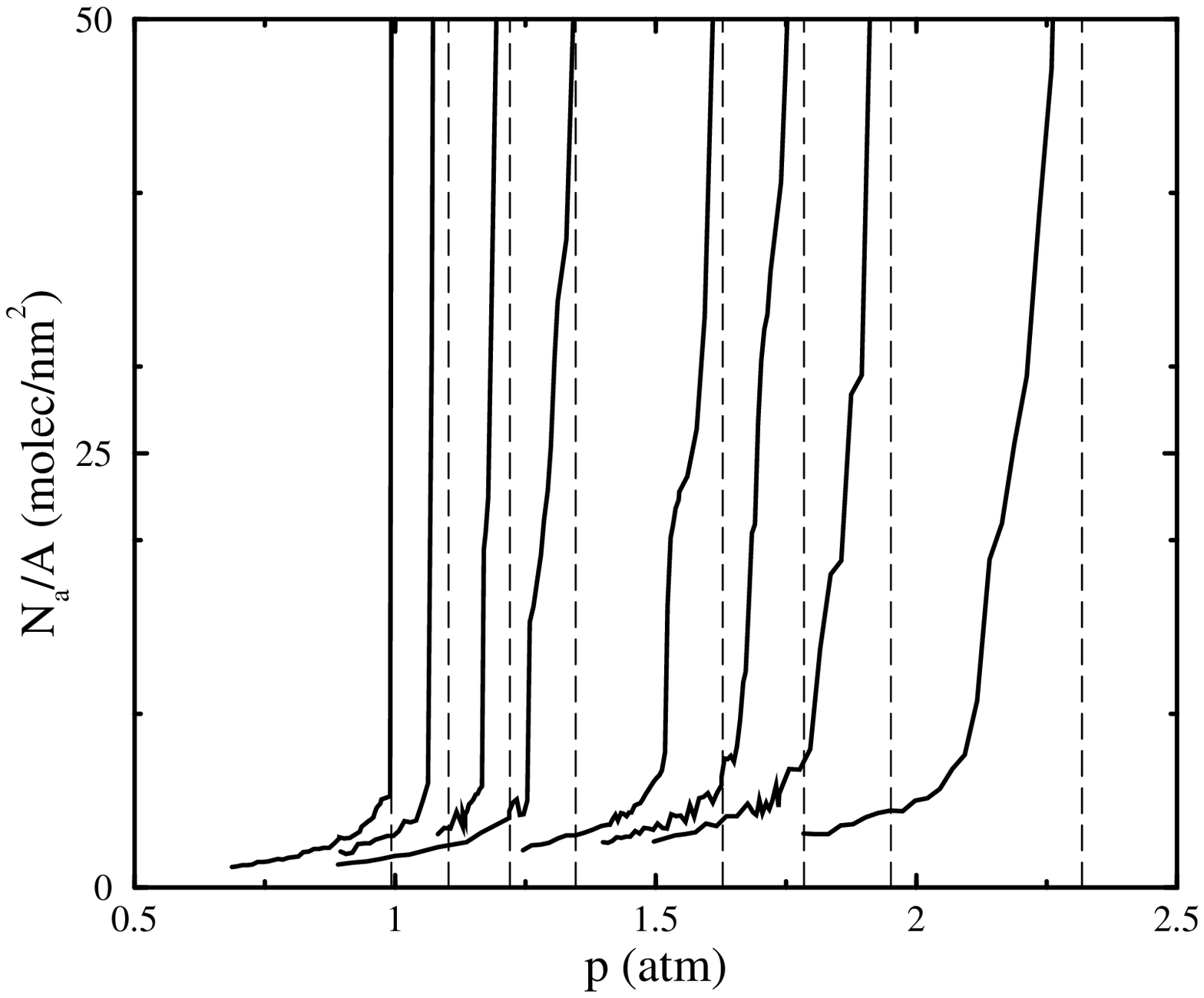}
\end{figure}
\vspace{5cm}\begin{center}{\bf FIG. 5}\end{center}

\newpage
\begin{figure}[ht]%%%%%%%%%%%% FIGURE 6
\epsfysize=5.in \epsfbox{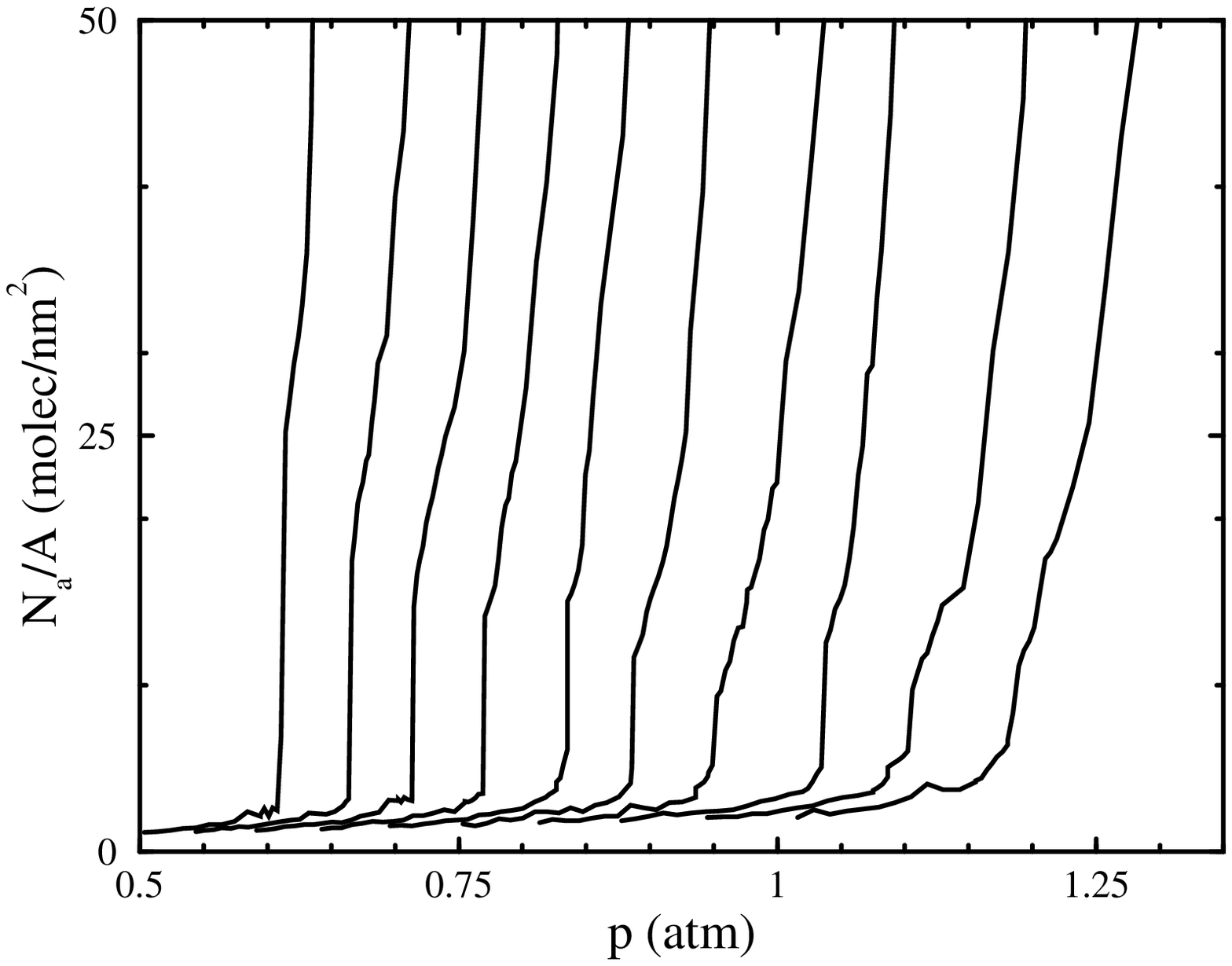}
\end{figure}
\vspace{4cm}\begin{center}{\bf FIG. 6}\end{center}

\newpage
\begin{figure}[ht]%%%%%%%%%%%% FIGURE 7
\epsfysize=5.in \epsfbox{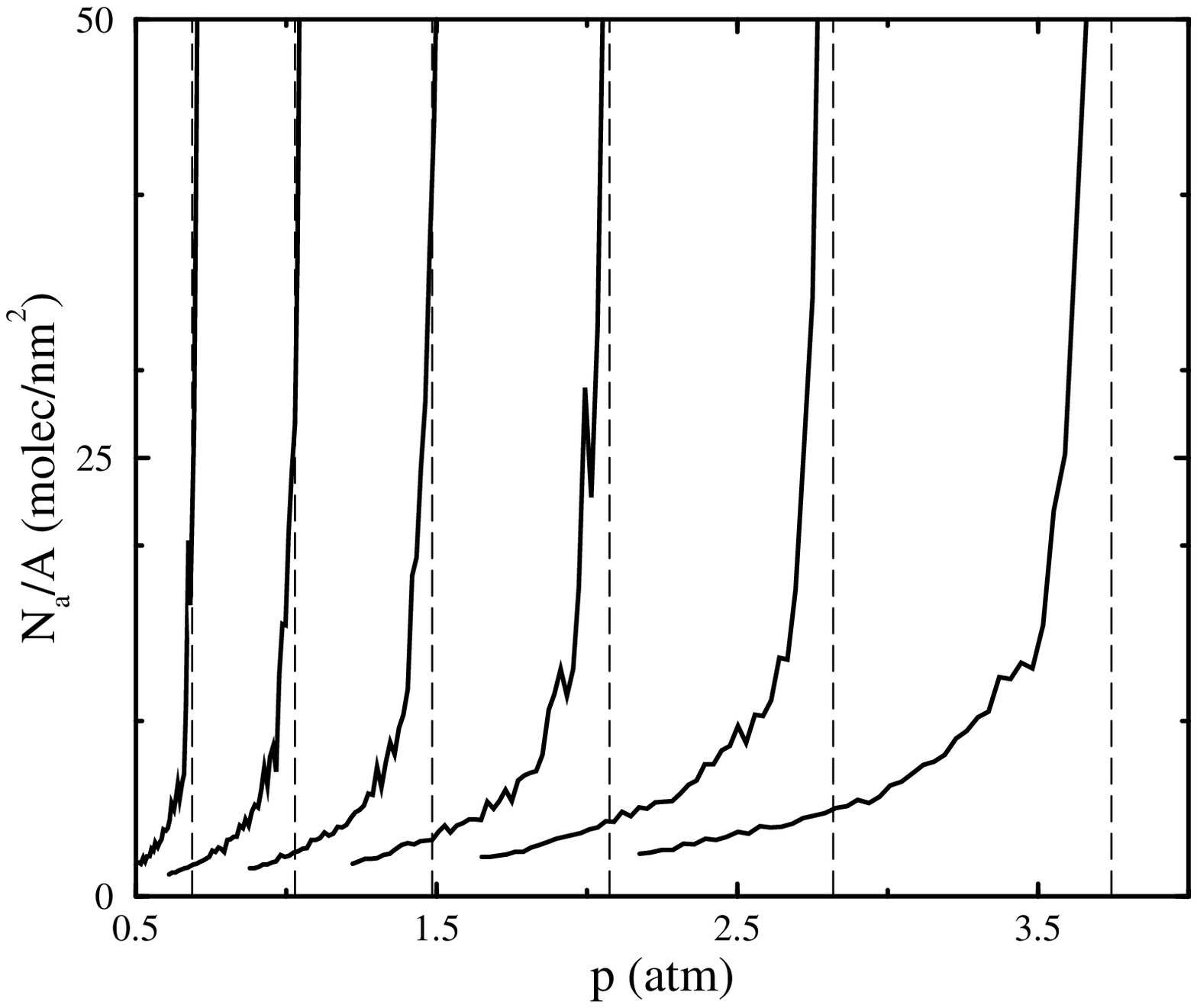}
\end{figure}
\vspace{5cm}\begin{center}{\bf FIG. 7}\end{center}

\newpage
\begin{figure}[ht]%%%%%%%%%%%% FIGURE 8
\epsfysize=5.in \epsfbox{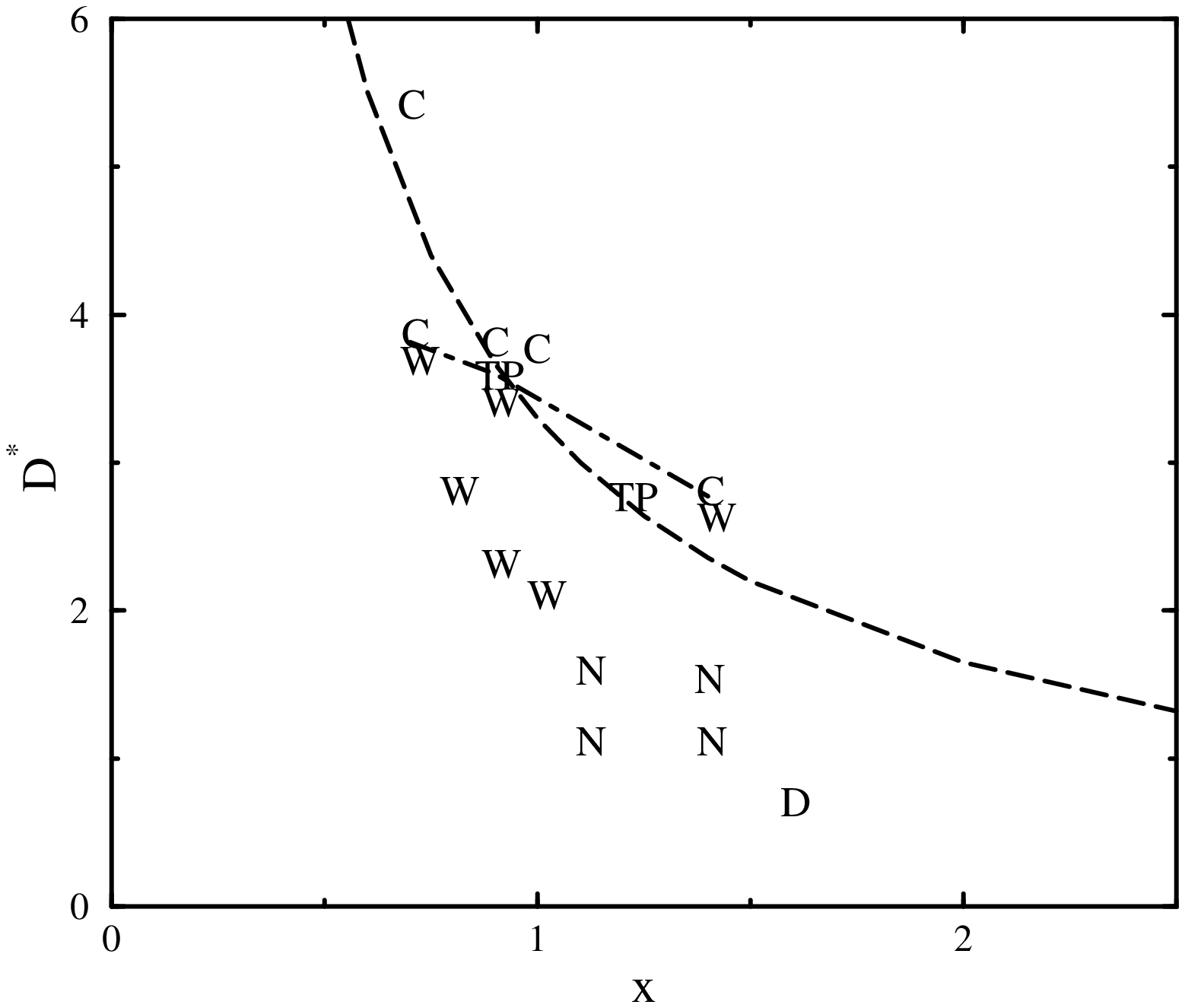}
\end{figure}
\vspace{4cm}\begin{center}{\bf FIG. 8}\end{center}


\begin{thebibliography}{99}
\bibitem{ref.3.pseudoMg}
The alert reader might notice that the adsorption potential 
parameters for Kr/PMg coincide with those for Xe/Li; this reveals the 
accidental origin of this serendipitous choice of fictitious substrate.
\end{thebibliography}
\end{document}